\documentclass[intlimits,twoside,a4paper]{article}

\usepackage[cp1251]{inputenc}
\usepackage[eqsecnum]{cmpj3}

\usepackage{bm}

\issue{2019}{22}{3}{33704}
\doinumber{10.5488/CMP.22.33704}

\title[BFH in the truncated Hilbert space limit]%
{Bose-Fermi-Hubbard model in the truncated Hilbert space limit}
\author[V.O. Krasnov, \framebox{I.V. Stasyuk}]{V.O. Krasnov, \framebox{I.V. Stasyuk}}
\address{
	Institute for Condensed Matter Physics of the National Academy of Sciences of Ukraine, 1 Svientsitskii St., 79011 Lviv, Ukraine
}
\date{Received	August 5, 2019, in final form August 28, 2019}

\begin{document}

\maketitle

\begin{abstract}
It is shown that in the Bose-Fermi-Hubbard model which is used for a description of the ultracold atomic boson-fermion mixture in the optical lattice, the $n_\text{B}\leqslant 2$ restriction enables one to analyze a more general case of separated lobes Mott insulator in comparison with the case of $n_\text{B} \leqslant 1$ (hard-core bosons). It also showed that the restriction to no more than 2 bosons on site is enough to comprehend the structure of $(\mu, |t_0|)$ diagrams at arbitrary $n^\text{b}$ values with an account of a possibility of the 1st order phase transition.
\keywords Bose-Fermi-Hubbard model, Bose-Einstein condensate, 
optical lattices, phase transitions 
\pacs 71.10 Fd, 71.38
\end{abstract}

\section{Introduction}
\vspace{-2mm}
We investigate the phase transitions in the Bose-Fermi-Hubbard model proposed for a description of thermodynamics and energy spectrum of the ultracold mixtures of Bose- and Fermi-atoms in optical lattices. Our main task is to study the influence of the fermion subsystem on the phase transition (PT) from the normal (Mott insulator, MI) phase to the superfluid (SF) phase with the Bose-Einstein condensate. Consideration is performed for the case of infinitely small fermion transfer in the limit of truncated basis of the single-site states of bosons ($n_i \leqslant 1$ for hard-core bosons, and $n_i \leqslant 2$). The boson-fermion interaction $U'$ is taken into account exactly while the hopping $t_0$ of bosons is considered within the mean-field approximation. 

The regime of fixed values of chemical potentials ($\mu$ and $\mu'$, respectively) of Bose- and Fermi-particles is the basic one in our study. Analyzing the behavior of the BE condensate order parameter and the grand canonical potential, we have built, for the case of hard-core bosons and ``heavy'' fermions, the $(\mu, \mu')$ and
$(\mu, t_0)$ phase diagrams at $T=0$ and at nonzero temperatures \cite{stasyuk:43702}. It is shown that: i) in the cases when transition to the SF phase is accompanied by the change of the mean number of fermions, the PT order changes from the 2nd to the 1st one; this result corresponds to the literature data (DMRG calculations for harmonic trap \cite{snoek:155301}); ii) the shape of the MI phase regions (so-called lobes) in the $(\mu,t_0)$ diagrams and the localization of the 1st order PT segments upon them depend on the $\mu'$ level position.

At the same time, in the literature \cite{panov:2017,szabados:208}, the approximation is developed, where the models are based on  truncated  Hilbert space with a larger number of the boson states. As an example, one can mention the  known approach, where $N_\text{max}^\text{B}=2,   (n_i^\text{b}=0,1, \text{and}~2)$ \cite{pekker:144527}.
\vspace{-2mm}

\section{Bose-Fermi-Hubbard model}
\vspace{-2mm}
In our investigation we use the Hamiltonian of the BFH model written in the form
\begin{align}
\label{eq:2.1}
H=-\sum\limits_{<i,j>} t_{ij}b^{+}_{i}b_{j}+\frac{U}{2}\sum\limits_{i}n_i^\text{b}(n_i^\text{b}-1)+U'\sum\limits_{i}n_{i}^\text{b}n_{i}^\text{f}-\mu'\sum\limits_{i}n_{i}^\text{f}-\mu\sum\limits_{i}n_{i}^\text{b}.
\end{align}
Here, $U$ and $U'$ are constants of boson-boson and boson-fermion on-site interactions; the chemical potentials of bosons and fermions are $\mu$ and $\mu'$, respectively, and $t$ is the tunneling amplitude of bosons describing the boson hopping between the nearest lattice sites. We restrict ourselves to the so-called case of ``heavy'' fermion when the hoping of fermions can be neglected.

 The occupation numbers of bosons (fermions) on the site $i$, are $n_{i}^\text{b}(n_{i}^\text{f})$ and we can write (as it was introduced in \cite{stasyuk:13003}) the single-site basis of states
\begin{flalign}
\label{eq:2.2}
&(n_{i}^\text{b}=n;n_{i}^\text{f}=0)\equiv|n,i\rangle; \quad (n_{i}^\text{b}=n;n_{i}^\text{f}=1)\equiv|\widetilde{n},i\rangle.
\end{flalign}

Then, the Hubbard operators are:
$X_{i}^{n,m}=|n,i\rangle\langle m,i|, X_{i}^{\widetilde{n},\widetilde{m}}=|\widetilde{n},i\rangle\langle \widetilde{m},i|$, etc.

The creation and annihilation operators and the opccupation number operators will be expressed in terms of $X$-operators in the following way:
\begin{flalign}
\label{eq:2.3}
&b_i=\sum\limits_n\sqrt{n+1}X_i^{n,n+1}+\sum\limits_{\widetilde{n}}\sqrt{\widetilde{n}+1}X_i^{\widetilde{n},\widetilde{n}+1},\nonumber\\
&b^+_i=\sum\limits_n\sqrt{n+1}X_i^{n+1,n}+\sum\limits_{\widetilde{n}}\sqrt{\widetilde{n}+1}X_i^{\widetilde{n}+1,\widetilde{n}},\\
&n_i^\text{b}=\sum\limits_n nX^{n,n}+\sum\limits_{\widetilde{n}} \widetilde{n}X^{\widetilde{n},\widetilde{n}},\quad n_i^\text{f}=\sum\limits_{\widetilde{n}} X^{\widetilde{n},\widetilde{n}}.\nonumber
\end{flalign}
The Hamiltonian in this new representation takes the form:
\begin{flalign}
\label{eq:2.4}
&H=H_0+H^\text{b}\,,\\
&H_0=\sum\limits_{i,n}\lambda_{n}X_i^{nn}+\sum\limits_{i,\widetilde{n}}\lambda_{\widetilde{n}}X_i^{\widetilde{n}\widetilde{n}}, \quad H^\text{b}=-\sum\limits_{<i,j>} t_{ij}b^{+}_{i}b_{j}\,,\nonumber\\
&\lambda_{n}=\frac{U}{2}n(n-1)-n\mu, \quad  \lambda_{\widetilde{n}}=\frac{U}{2}\widetilde{n}(\widetilde{n}-1)-\mu\widetilde{n}-\mu'+U'\widetilde{n}.\nonumber
\end{flalign}
Using the order parameter of BE condensate $\varphi=\langle b_i\rangle = \langle b_i^+\rangle$ we can write in the case of mean field approximation (MFA) \cite{stasyuk:13003}:
 \begin{align}
 \label{eq:2.5}
 &b_i^+b_j\rightarrow \varphi(b_i^++b_i)-\varphi^2 \\
 &\sum\limits_{ij}t_{ij}b_i^+b_j=\varphi t_0\sum\limits_i(b_i^++b_i)-Nt_0\varphi^2,\nonumber
 \end{align}
 (here, $t_0=\sum t_{ij}=-|t_0|, t_0<0$).
 
 Then, for initial Hamiltonian after separating the mean field part we will have:
 \begin{align}
 \label{eq:2.6}
 &H_\text{approx}=H_\text{MF}+\sum\limits_{ij}t_{ij}(b_i^+-\varphi)(b_i-\varphi)\,,
 \end{align}
where
\begin{align}
 \label{eq:2.7}
&H_\text{MF}=\sum_i H_i-Nt_0\varphi^2, \quad H_i=\sum\limits_{pr}H_{pr}X^{pr}_i.
\end{align}
The matrix $H_{pr}$ is a block-diagonal matrix and is of the following form:
\begin{align}
\label{eq:2.8}
H_{pr}=\left(
\begin{array}{cc}
H_{pr}^{11} &  0  \\
0    &  H_{pr}^{22}
\end{array}
\right),
\end{align}
 where $H_{pr}^{11}$ is infinite matrix:
 \vspace{-2mm}
\begin{align}
\label{eq:2.9}
H_{pr}^{11}=\left(
\begin{matrix}
\lambda_{0}&  |t_0|\varphi  &   0 & 0 & \ldots  \\
|t_0|\varphi &  \lambda_{1} & \sqrt{2}|t_0|\varphi &0 & \ldots \\
0   & \sqrt{2}|t_0|\varphi & \lambda_{2}  & \sqrt{3}|t_0|\varphi & \ldots \\
0 & 0 & \sqrt{3}|t_0|\varphi & \lambda_{3} & \ldots \\
\ldots  &   \ldots       &\ldots & \ldots   &  \ldots   
\end{matrix}\right),
\end{align} 
and $H_{pr}^{22}$ is of the same form but with the replacement  $n\rightarrow \tilde{n}$ 
[see (\ref{eq:2.2})].
\newpage
This form of the Hamiltonian is widely used while investigating the thermodynamical properties of this type of models [Bose- Hubbard (BH) and Bose-Fermi Hubbard (BFH) models]
in the case of optical lattices. Discussing the applicability of the MFA in such models we have to state that this approximation is more suitable for systems with a great number of interacting sites connected with the given one. The size of the system should set up the thermodynamical limit  ($N\rightarrow\infty$) and in the case of experiments in optical lattices the number of particles is of the order of  $N\approx 10^4 \ldots 10^5$. As we see, it is quite sufficient to satisfy the macroscopicity condition.

\section{Spinodal lines and the role of boson transfer (an example: the ground state of BFH) }

In this section and in all of the rest, the calculations were performed for the ground state ($T=0$) case and for the lowest possible state to calculate the value of temperature $T=0.005U'$ (for the $n_\text{B} \leqslant 2$ case).

\begin{figure}[!b]
	\includegraphics[width=0.45\textwidth]{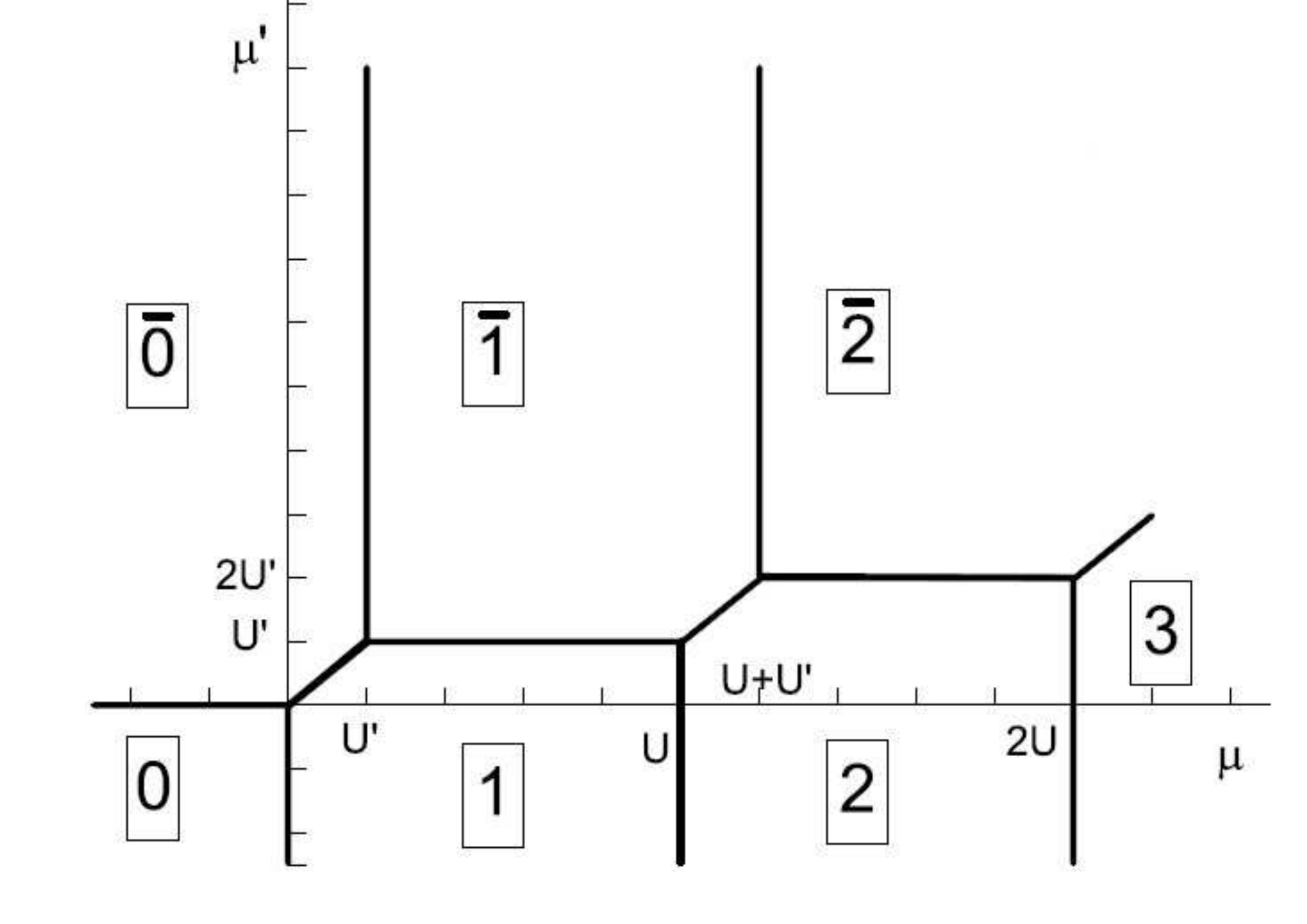}\hfill
	\includegraphics[width=0.4\textwidth]{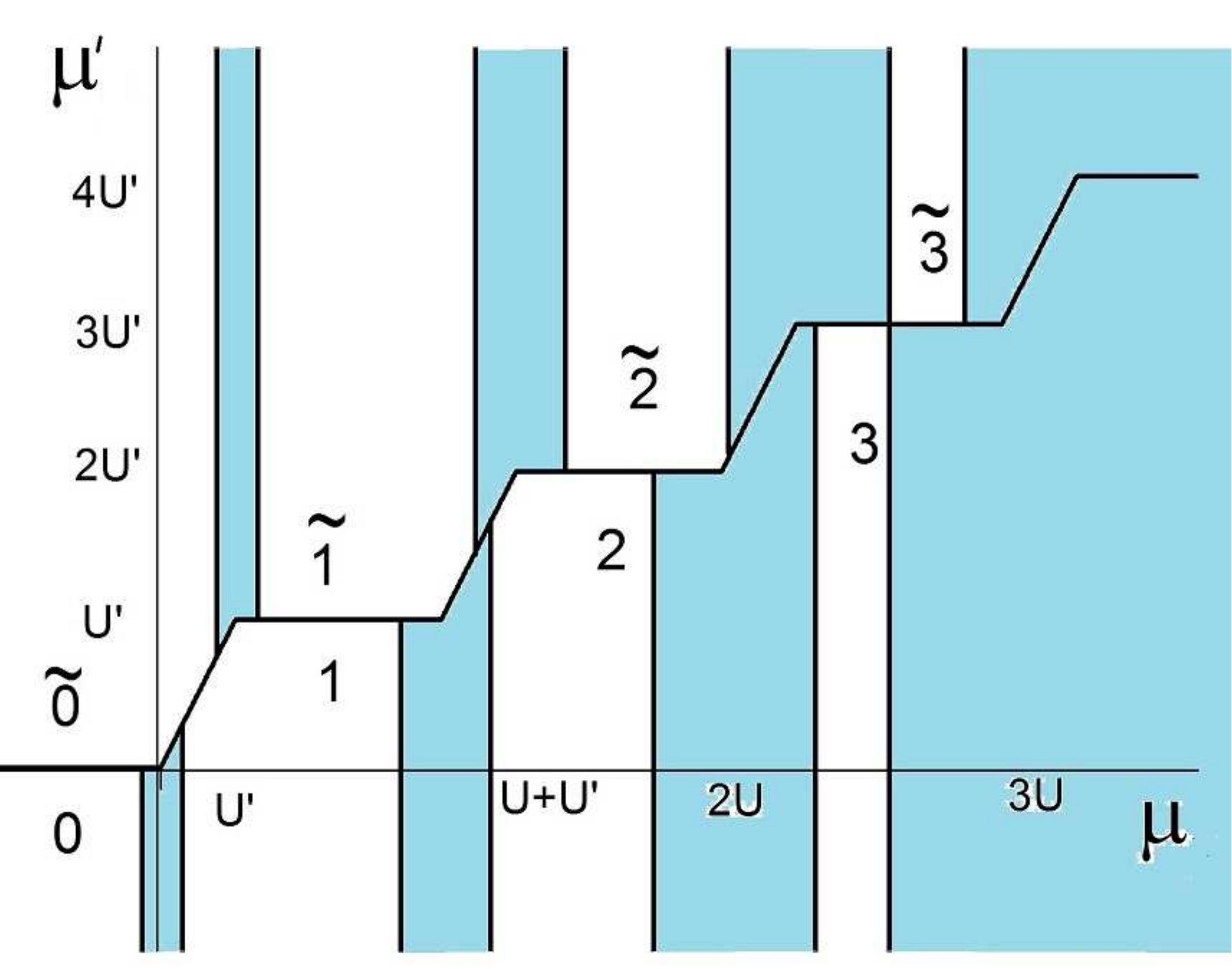}
	\caption{(Colour online) Ground state phase diagrams for BFH model at $|t_0|=0$ (on the left-hand side), the figure taken from \cite{stasyuk:13003}, and $|t_0|\neq 0$ (on the right-hand side).}
	\label{fig:fig1}
\end{figure}

Using the unperturbed Hamiltonian, we can build the ground state diagram for $(\mu, \mu')$ plane and get the spinodals (the lines of instability of the SF phase); at the same time, they are the PT lines in the case of the 2nd order transition. In the absence of boson transfer they split up the plane into different ground state regions \cite{stasyuk:13003} (figure~\ref{fig:fig1}, on the left). The vertical lines separates the states with a different number of bosons but with the same number of fermions while the slanting lines separate the states with different numbers of both bosons and fermions. A possibility of such transitions (with different numbers of both bosons and fermions) is the manifestation of the main differences between the BFH and pure BH models.

In the paper \cite{stasyuk:13003}, we described in what way using random phase approximation (RPA) for two-time Green's function $G_{ij}(t-t')=\langle\langle b|| b^+\rangle\rangle$ we can obtain the next equation for Green's function $G_{\bf k}(\omega)$
\begin{align}
\label{eq:3.1}
	G_{\bf k}(\omega)=\frac{1}{2\piup}\frac{g^{0}(\omega)}{1-g^{0}(\omega)t_{\bf k}}\,, \qquad  g^0(\omega)=\sum_{m}\left[ \frac{Q_m (m+1)} {\omega-\Delta_m}+\frac{Q_{\tilde{m}} (\tilde{m}+1)} {\omega-\Delta_{\tilde{m}}}\right],
\end{align}
where  $\Delta_{m}=\lambda_{m+1}-\lambda_{m}$ and  $ Q_{m}=\langle
X^{m,m}-X^{m+1,m+1}  \rangle$.

Instability connected with the phase transition from the Mott insulator to SF phase is characterized by the divergence of the Green's function $G_{{\bf {k}}=0}(\omega=0)\to \infty$, this condition leads to equation:
\begin{align}
\label{eq:3.2}
1= t_0\sum_{m}\left[ \frac{Q_m (m+1)} {\omega-\Delta_m}+\frac{Q_{\tilde{m}} (\tilde{m}+1)} {\omega-\Delta_{\tilde{m}}}\right]. 
\end{align}
Solving this equation (known as equation for spinodal lines) we can build the ground state diagram on the plane $(\mu, \mu')$ in the presence of the boson hopping (see figure~\ref{fig:fig1} on the right).

As we see, regarding these two phase diagrams, the boson hopping leads to the appearance of the SF phase in the regions where the number of bosons on site changes (painted areas). 

\begin{figure}[!t]
		\includegraphics[width=0.45\textwidth]{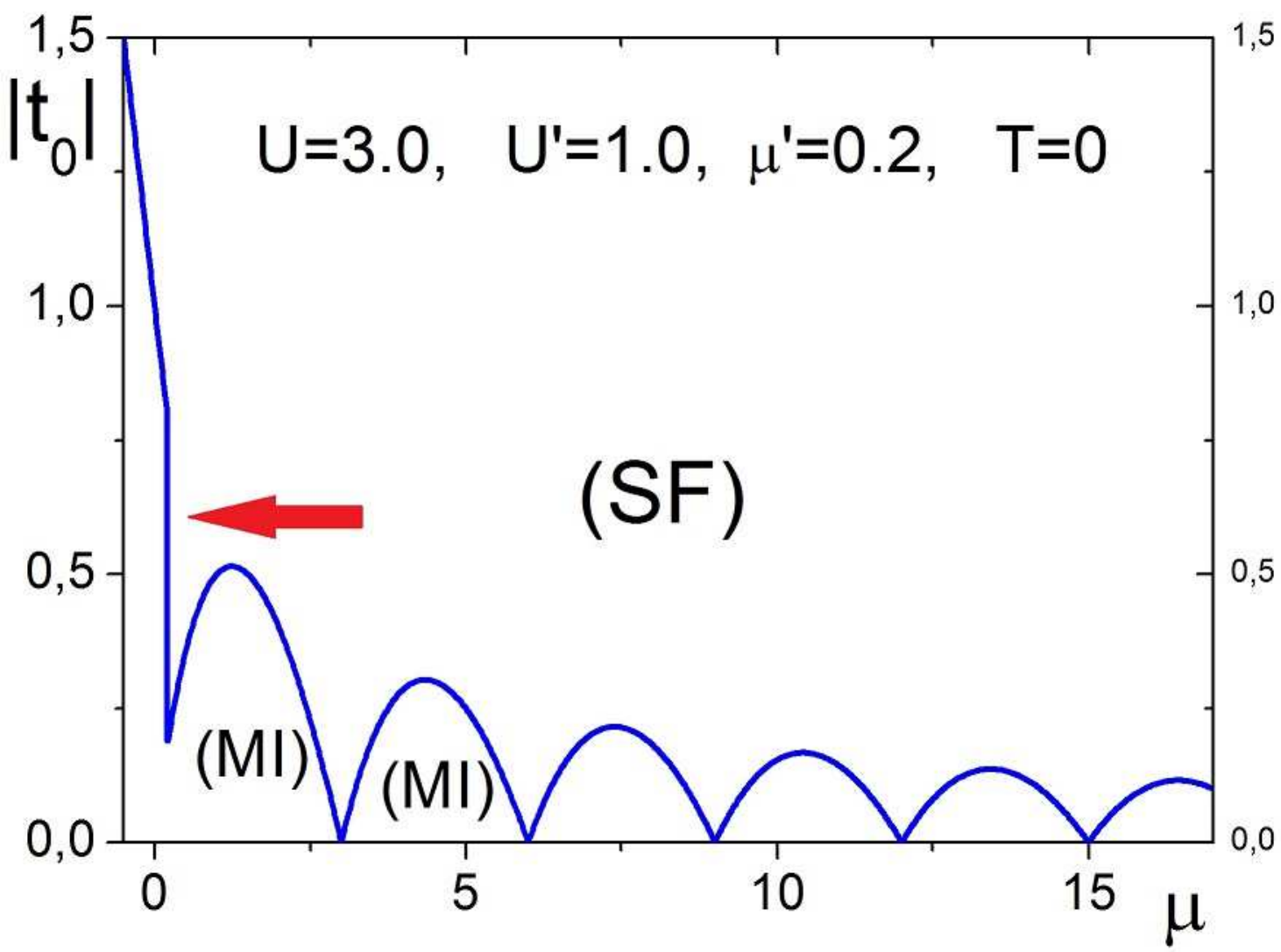}\hfill
			\includegraphics[width=0.45\textwidth]{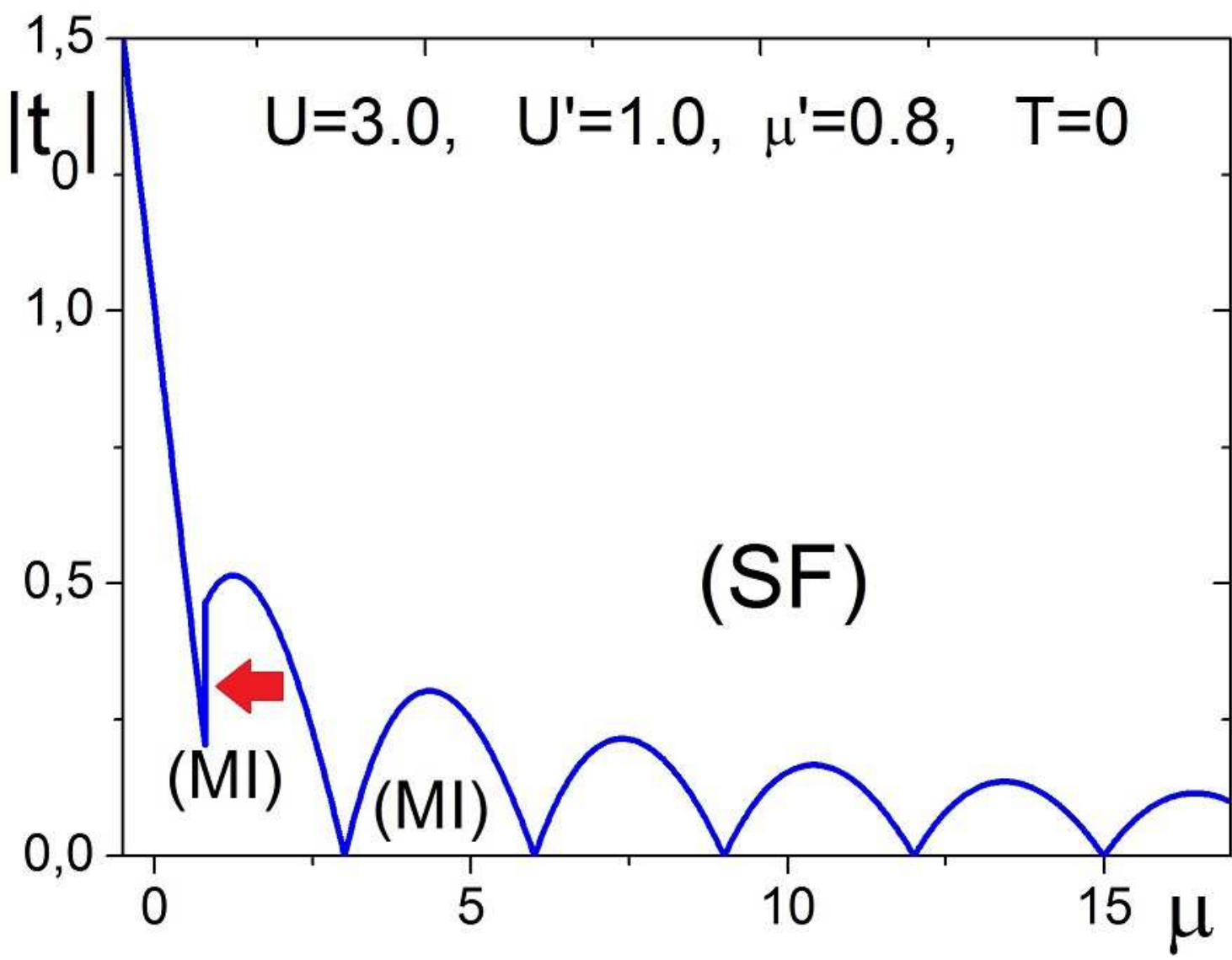} \\
			\includegraphics[width=0.45\textwidth]{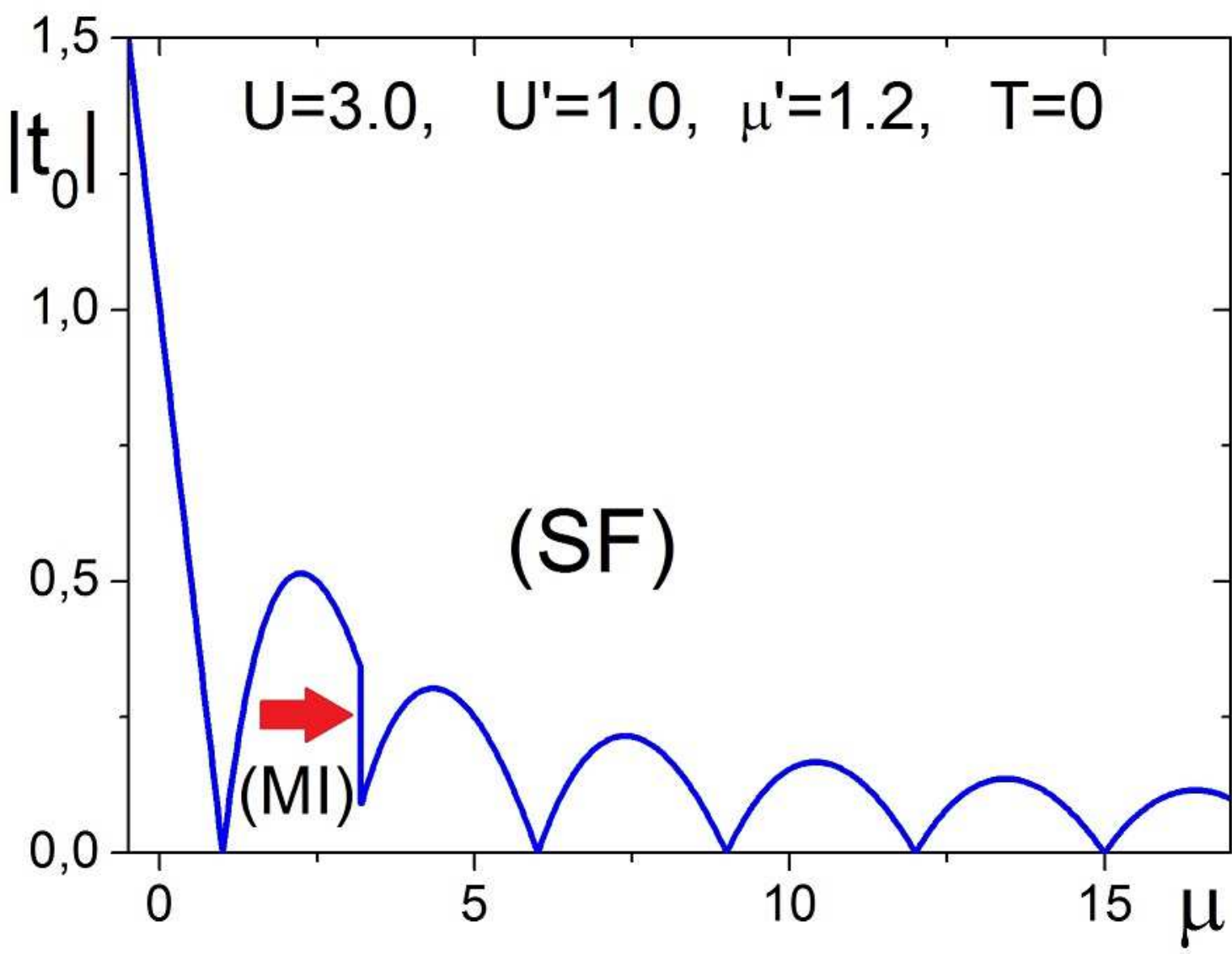}	\hfill
				\includegraphics[width=0.45\textwidth]{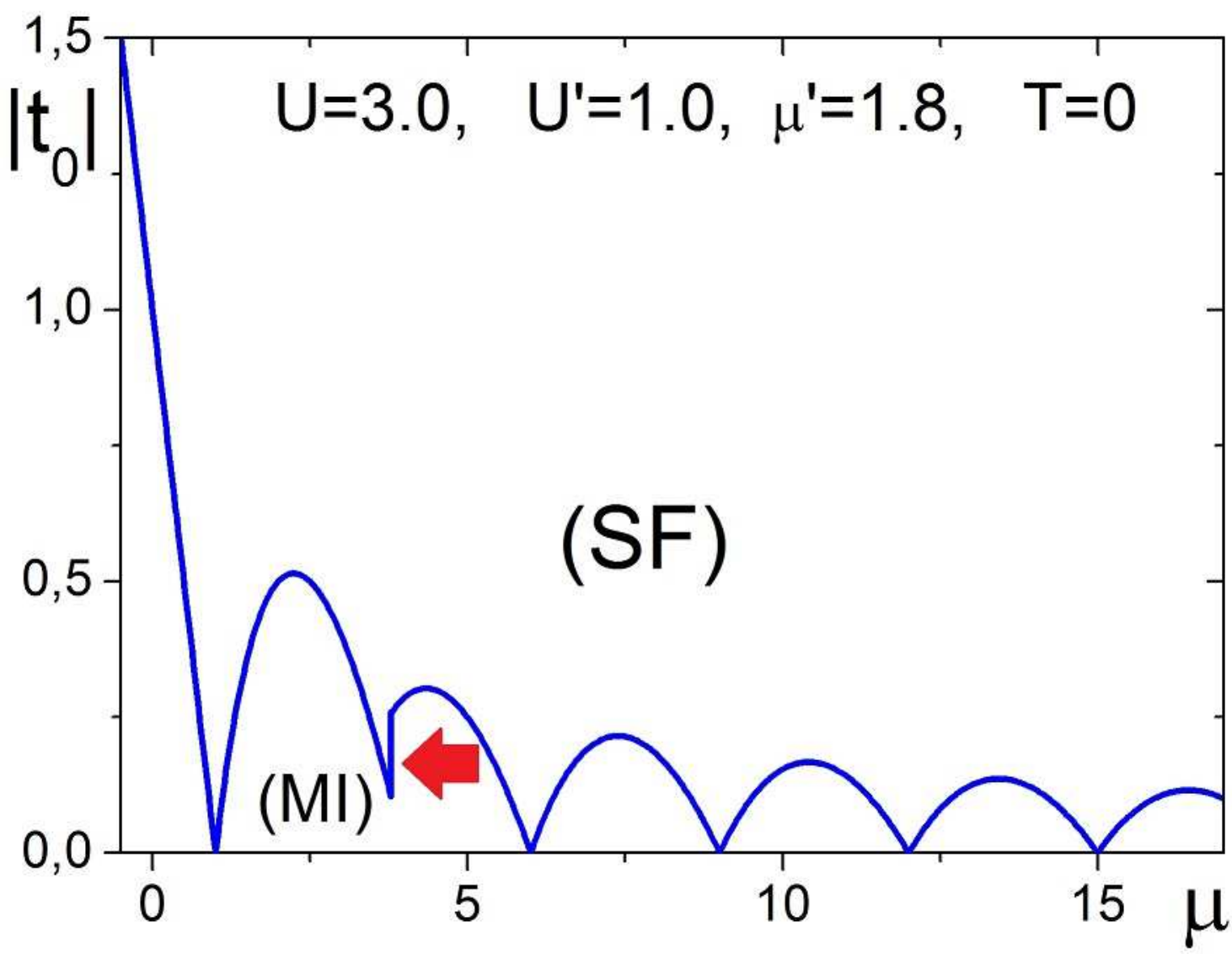}
				\caption{(Colour online) Spinodal lines on the $(\mu, |t_0|)$ plane at different values of chemical potential of fermions. Breaks of line are marked by red arrows.}
				\label{fig:fig2}
\end{figure}

Solving the equation~(\ref{eq:3.2}) for different values of chemical potential of fermions we will get the phase diagrams on the $(\mu, |t_0|)$ plane (figure~\ref{fig:fig2}). We see, similar  to the BH case, the sequence of lobes which separates the MI (under lobes) and SF phases. However, the presence of a fermion subsystem manifests itself here by the breaks on the separating line. 

\section{Truncated Hilbert space and first order phase transitions}
The above applied way of construction of spinodales is based on the investigation of  lines where the SF phase becomes unstable. A more complete analysis requires to search for the lines of co-existence of different phases, where the main criterium is an equality of thermodynamical potentials of neighbour phases (MI phase with $\varphi=0$ and SF phase with $\varphi\neq 0$).

Starting from equation~(\ref{eq:2.7}) we will obtain:
\begin{align}
\label{eq:4.1}
\Omega=-\Theta \ln \text{Sp}~\re^{-\beta H_\text{MF}}=-\Theta \ln\sum\limits_{\alpha}\re^{-\beta\varepsilon_{\alpha}}-Nt_0\varphi\,,
\end{align}
where 
\begin{align*}
H_\text{MF}=\sum_{i}\Big[H_i+ (b_i+b_i^+)\Big(\sum_{i}t_{ij}\varphi\Big)\Big]-Nt_0\varphi^2.
\end{align*}
In terms of Hubbard operators, the task, at the beginning, is reduced to diagonalization of the Hamiltonian matrix obtained after transition to the ``$n$'' -- representation
\begin{align}
\label{eq:4.2}
H_{pr}=\left(
\begin{array}{cc}
H_{pr}^{11} &  0  \\
0    &  H_{pr}^{22}
\end{array}
\right).
\end{align}
Here, the first infinite matrix is related to the states with zero fermions and the second one is related to the states with one fermion [see (~\ref{eq:2.2})].

Such an approach was used by Sheshandri \textsl{et al.} in \cite{sheshandri:257}, where the problem was solved by a numerical method restricted to the finite maximal number of the single-site boson states (the value of $N_\text{max}^\text{B}$  was determined by calculational ability).

At the same time, in the literature, an approximation is developed, where the models are based on truncated  Hilbert space with a larger number of the boson states \cite{panov:2017,szabados:208,pekker:144527}. As an example, one can mention the  known approach, where $N_\text{max}^\text{B}=2,   (n_i^\text{b}=0,1, \text{and}~2)$.

The case $N_\text{max}^\text{B}=2,   (n_i^\text{b}=0,1, \text{and}~2)$ opens up new possibilities. Let us stop on these more in detail compared with the hard-core boson case.

1. For the hard-core bosons, the matrix is of the form
	\begin{align}
	\label{eq:4.3}
	H_{pr}=\left(
	\begin{matrix}
	\lambda_{0}&  |t_0|\varphi  &   0 & 0   \\
	|t_0|\varphi &  \lambda_{1} & 0  & 0  \\
	0   & 0 & \lambda_{\tilde{0}}  & |t_0|\varphi  \\
	0 & 0 & |t_0|\varphi & \lambda_{\tilde{1}} 
	\end{matrix}\right),
	\end{align} 
where separate $(2\times2)$ blocks correspond to the states with $n_i^\text{f}=0$  and $n_i^\text{f}=1$ , respectively. Its eigenvalues are equal to (see ~\cite{stasyuk:43702}):
\begin{align}
\label{eq:4.4}
\varepsilon_{0',1'}=-\frac{\mu}{2}\pm\sqrt{\frac{\mu^2}{4}+t_0^2 \varphi^2}\,, \quad \varepsilon_{\tilde{0}',\tilde{1}'}=-\mu'-\frac{\mu}{2}+\frac{U'}{2}\pm\sqrt{\frac{(\mu-U')^2}{4}+t_0^2 \varphi^2}
\end{align}
and depend on the BE-condensate order parameter. Minimization of $\Omega(\varphi)$ with the help of equation $\partial\Omega/\partial \varphi=0$ determines the stable equilibrium  states of the system. At $T=0$, the problem can be solved analytically \cite{stasyuk:43702}.

2. Three-state model $(n_i^\text{b}=0,1,2)$.

Now, the matrix $H_{pr}$ is written in the following form:
\begin{align}
\label{eq:4.5}
H_{pr}=\left(
\begin{matrix}
\lambda_{0}&  |t_0|\varphi  &   0 & 0  &0 & 0  \\
|t_0|\varphi &  \lambda_{1} & \sqrt{2}|t_0|\varphi  & 0 &0 &0  \\
0   &  \sqrt{2}|t_0|\varphi  & \lambda_{2}  &  0& 0& 0 \\
0 & 0 & 0& \lambda_{\tilde{0}} & |t_0|\varphi & 0 \\
0 & 0 & 0& |t_0|\varphi & \lambda_{\tilde{1}} & \sqrt{2}|t_0|\varphi  \\
0 & 0 & 0& 0 & \sqrt{2}|t_0|\varphi & \lambda_{\tilde{2}}  	
\end{matrix}\right).
\end{align} 
Equation for eigenvalues
\begin{align}
\label{eq:4.6}
\text{det}~||H_{pr}-\varepsilon\delta_{pr}||=0\,,
\end{align}
 can be easily written.

However, in spite of significant simplification of the model, which is a consequence of truncation of the basis of states, the problem is further solved in a numerical way [although here there exists a possibility of the analytical solution for $(3\times3)$ case].

It is important to stress that the truncated basis approach in the above described form, can be applied to the cases of a sequence of any of the three neighbor boson states $(n-1,n,n+1)$. The same situation takes place in the case of hard-core bosons and concerns the sequences of two neighbor states [$(0,1)$ or $(1,2)$, $(2,3)$ etc.; in general, $(n-1,n)$].

\subsection{$(\mu,|t_0|)$ phase diagrams}
	
Solving the equation~(\ref{eq:4.6}) we numerically get the eigenvalues  $\varepsilon_{n'}$ which can be used in constructing the diagonal Hamiltonian in the form:
\begin{align*}
H=\sum_{i,p'}\varepsilon_{p'}X_i^{p'p'}-t_0\varphi^2,
\end{align*}	
thus, as the next steps, there follows an equation for $\varphi$ by minimizing $\Omega$ and finding the values of the order parameter. The stable states and lines of the phase transition of the 1st order can be found from the global minimum condition for the grand canonical potential.

Using the obtained 1st order PT lines and  equations for spinodals we have a possibility to build phase diagrams on the plane $(\mu,|t_0|)$.

\begin{figure}[!b]
	\includegraphics[width=0.45\textwidth]{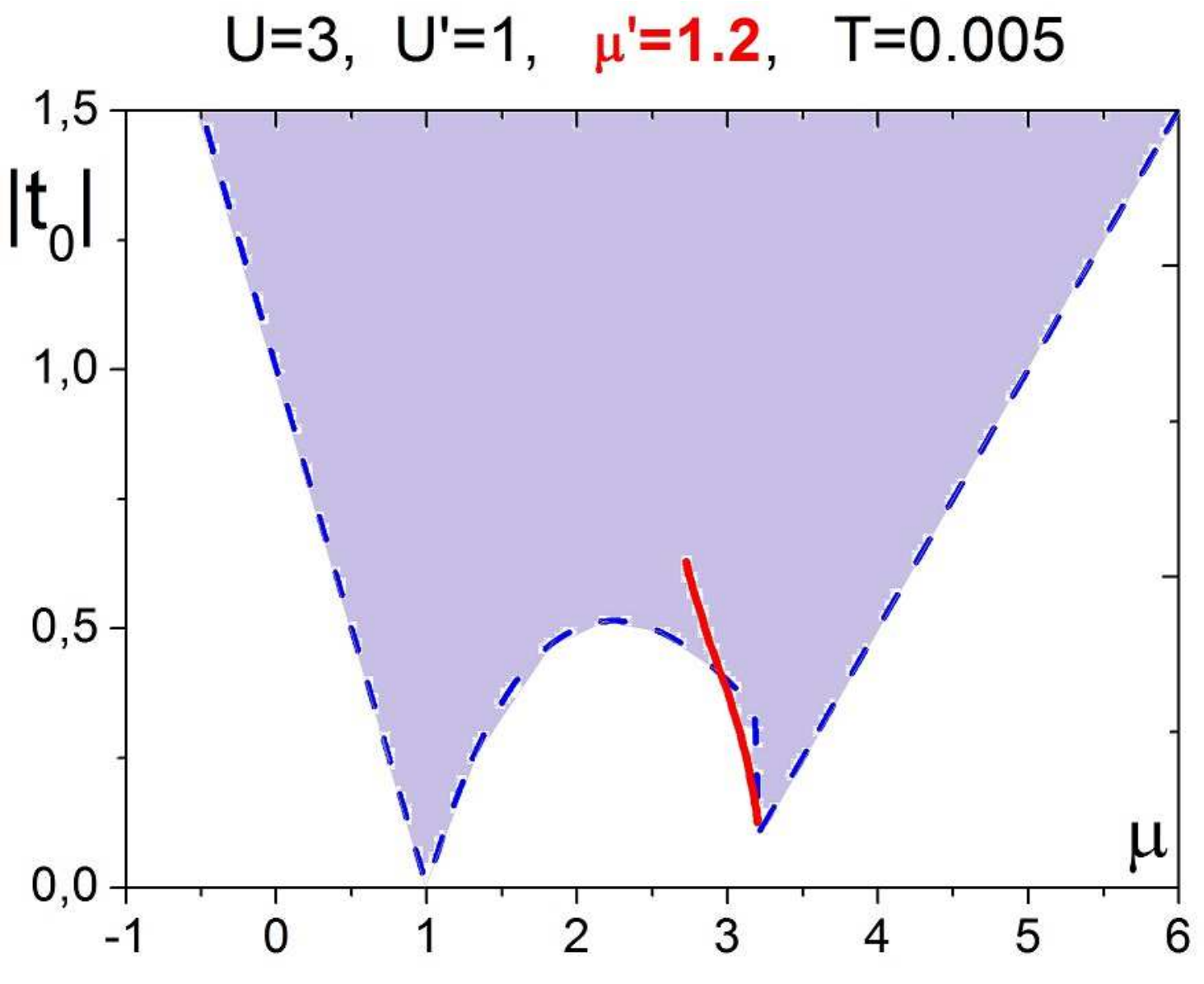}\hfill
	\includegraphics[width=0.45\textwidth]{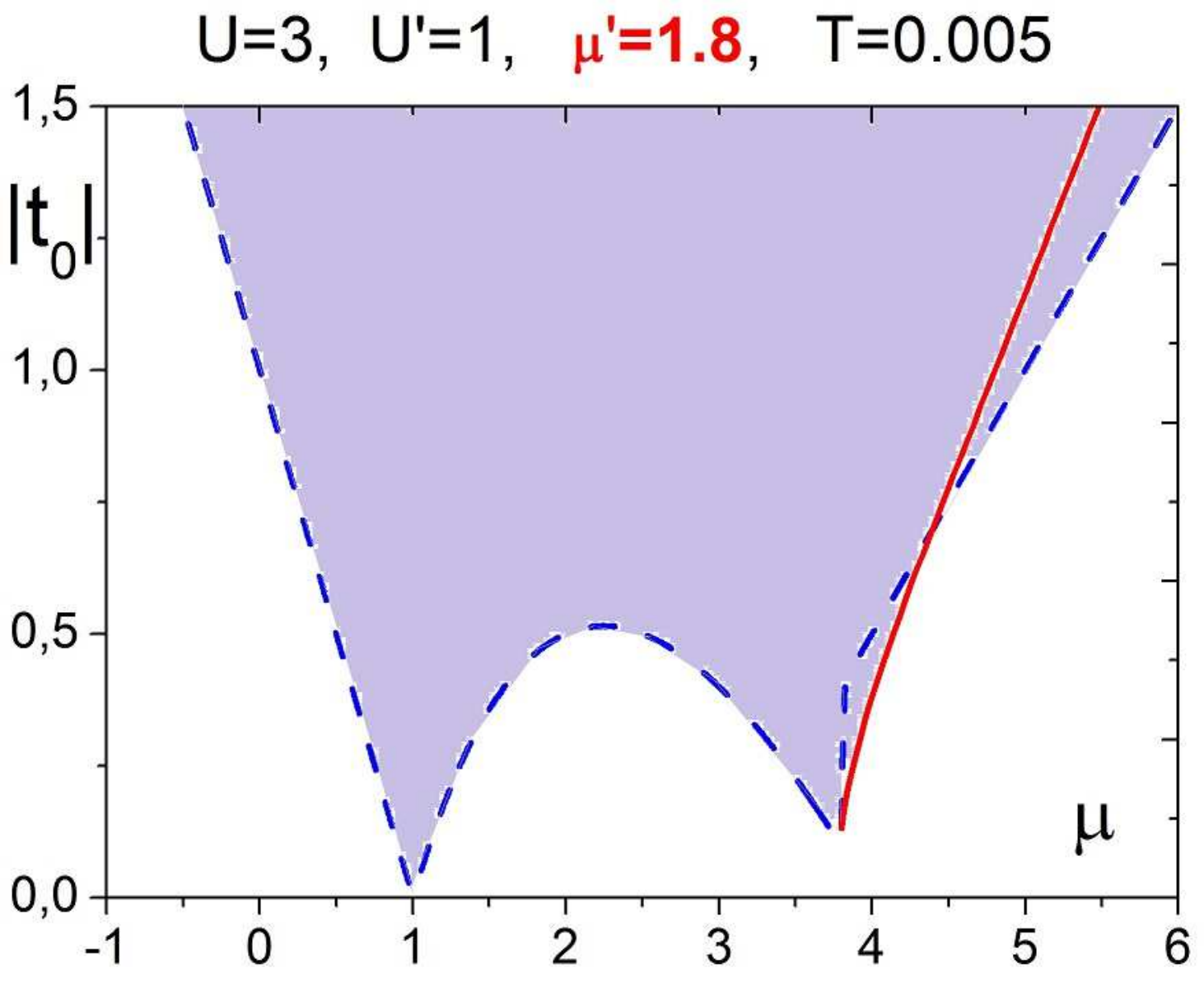}
	\caption{(Colour online) $(\mu,|t_0|)$ phase diagrams for different chemical potential values of fermions. Here and hereafter, the curves for 1st (2nd) order PT are depicted as solid (dashed) lines.}
	\label{fig:fig3}
\end{figure}	
As one can see from figure~\ref{fig:fig3}, an increase of the chemical potential of fermions leads to a rebuilding of the lobe of MI-SF phase transition and to the change of the 1st order PT position.  

\subsection{Comparing different reductions of Hilbert space}

\begin{figure}[!t]
	\includegraphics[width=0.45\textwidth]{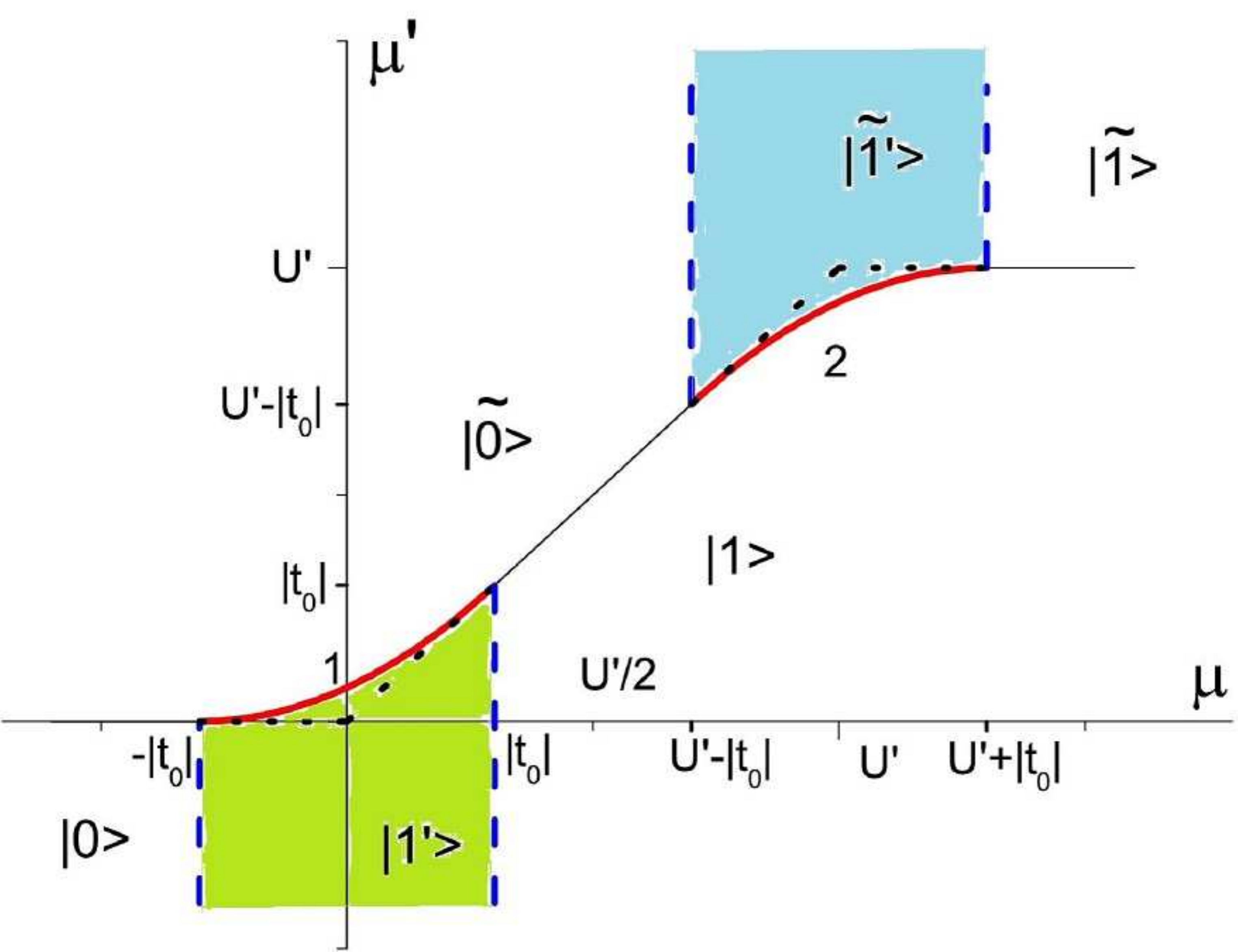}\hfill
	\includegraphics[width=0.45\textwidth]{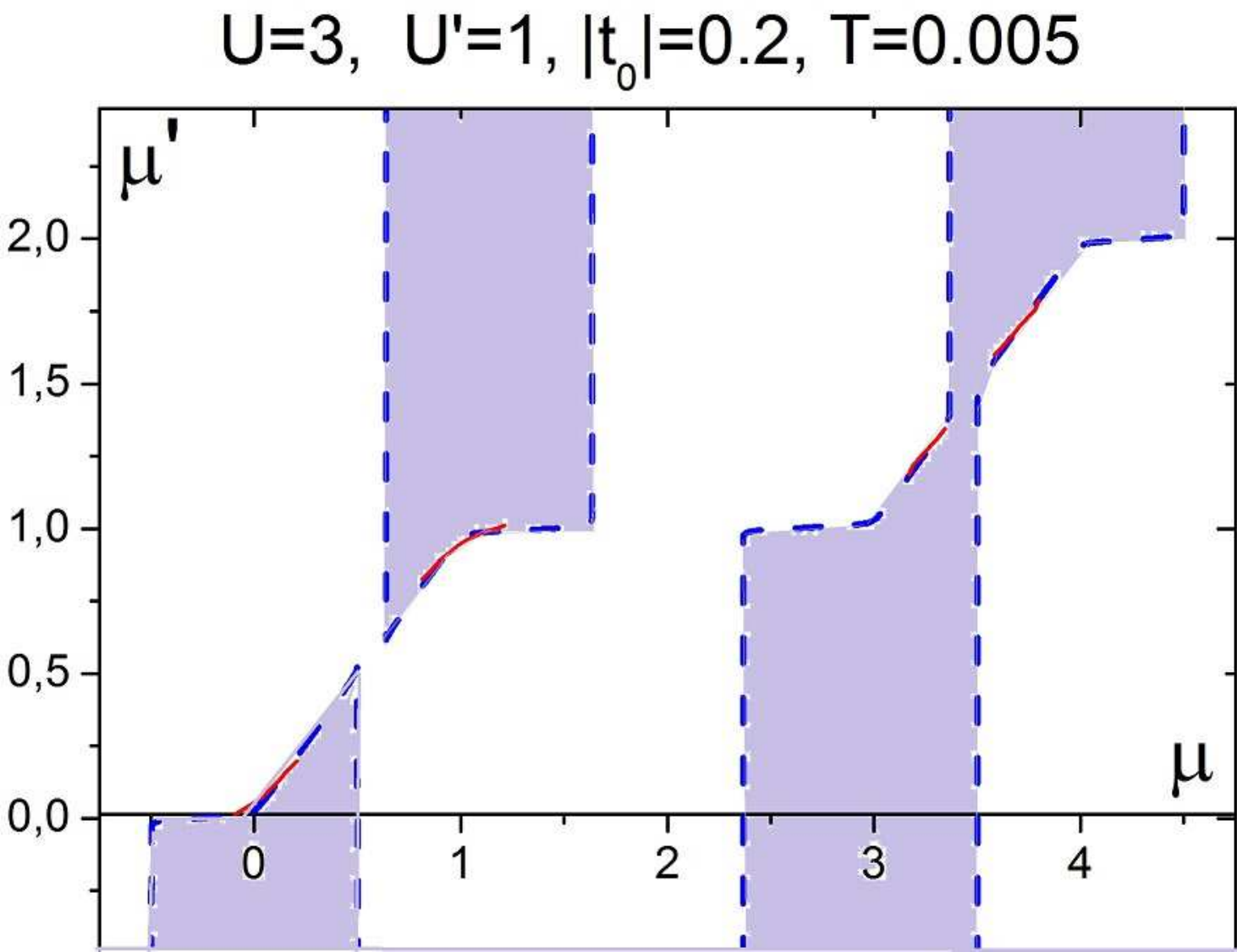}
	\caption{(Colour online) $(\mu, \mu')$  phase diagrams. On the left-hand side: hard-core bosons, ground state diagram for $n_\text{B} \leqslant 1$ is taken from \cite{stasyuk:43702};  the diagram for $n_\text{B}\leqslant 2$ case (on the right-hand side); $T=0.005U'$.}
	\label{fig:fig4}
\end{figure}

\begin{figure}[!t]
	\includegraphics[width=0.4\textwidth]{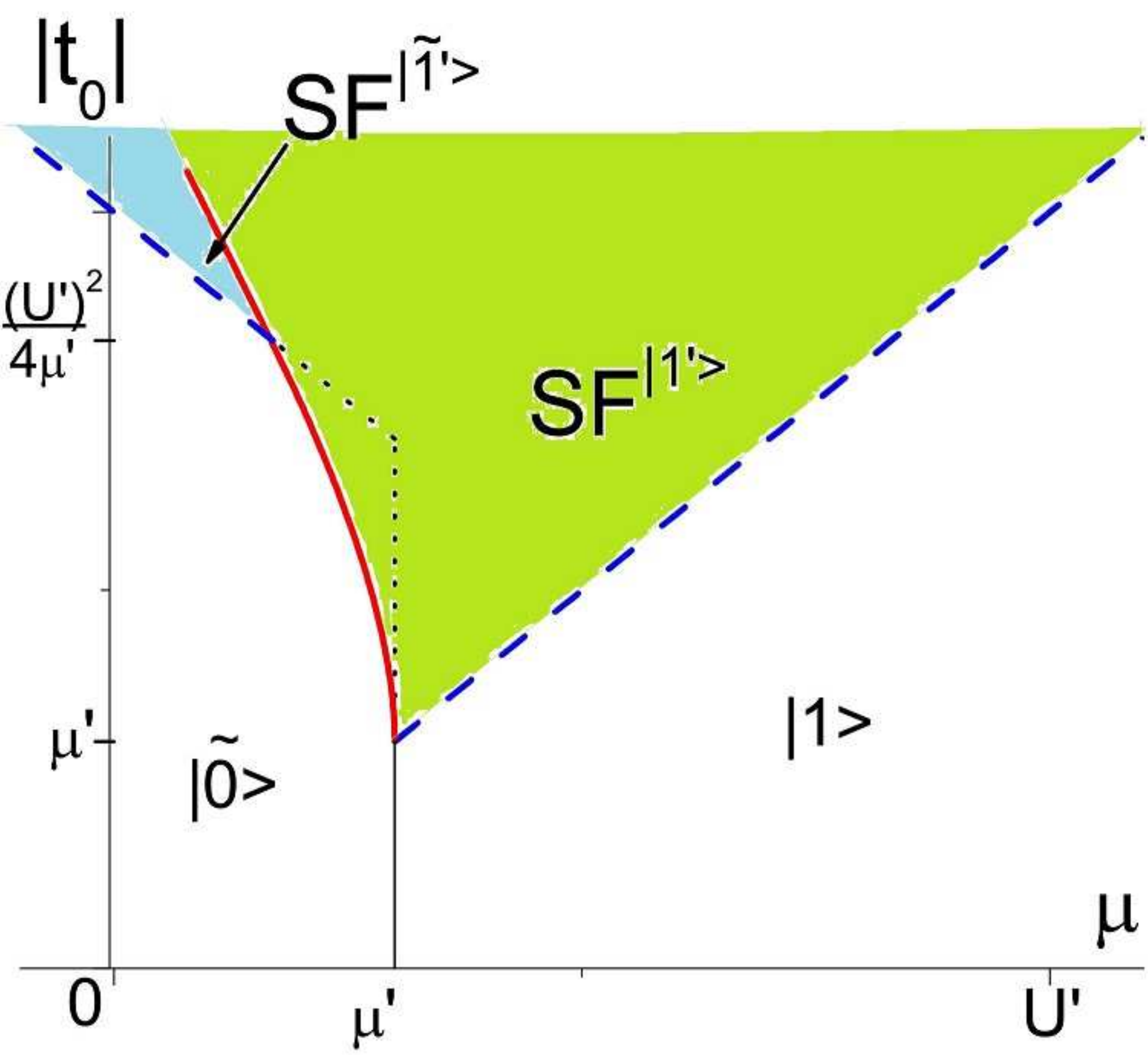}\hfill
	\includegraphics[width=0.45\textwidth]{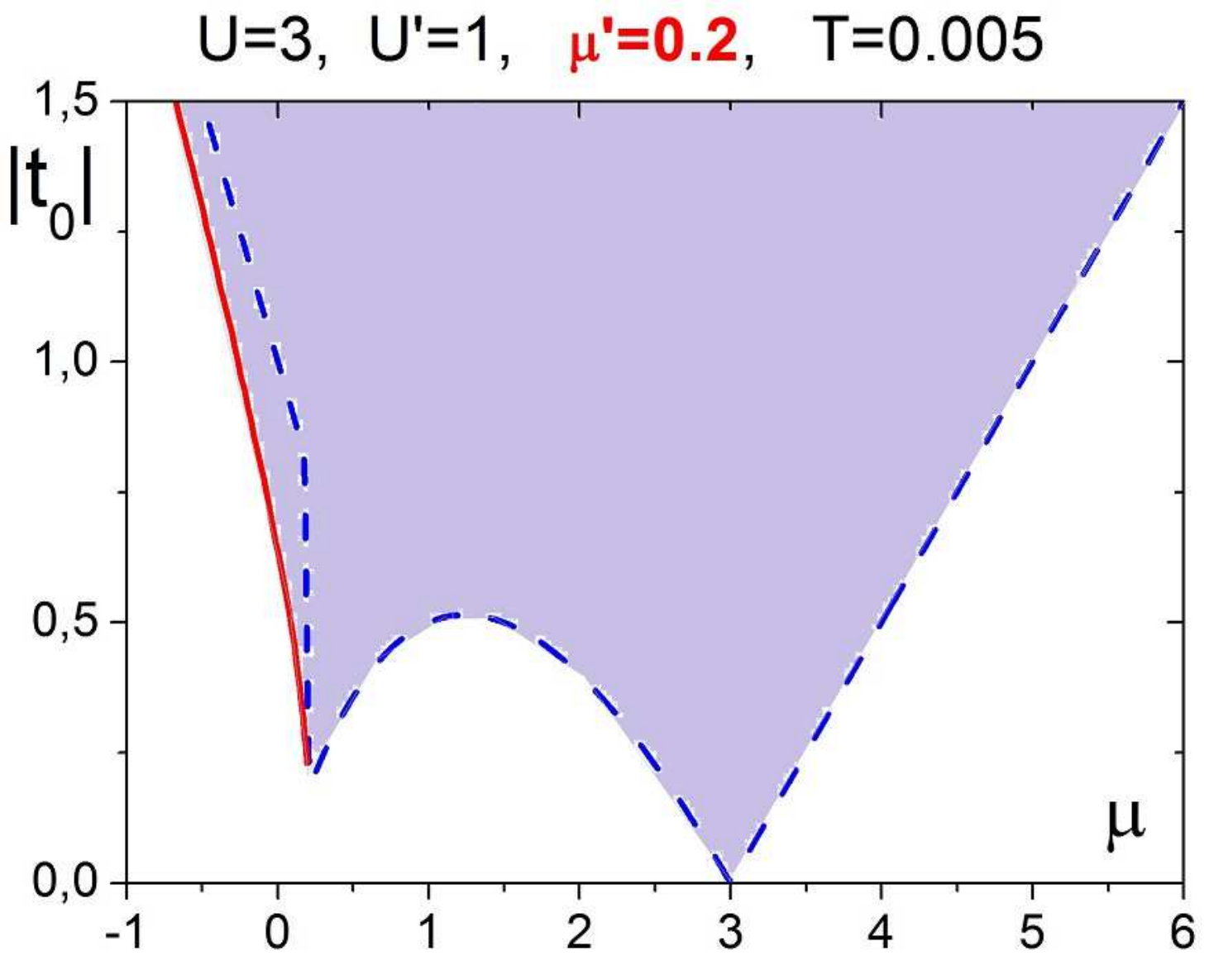}
	\caption{(Colour online) $(\mu, |t_0|)$  phase diagrams. Ground state for $n_\text{B} \leqslant 1$ (on the left-hand side) taken from \cite{stasyuk:43702}. $T=0.005U'$ diagram for the $n_\text{B} \leqslant 2$ (on the right-hand side).}
	\label{fig:fig5}
\end{figure}

In order to compare different (with different maximal  number of bosons on site) approximations we can use the earlier obtained (see \cite{stasyuk:43702})  for the case of $n_\text{B}\leqslant 1$ (hard-core bosons) phase diagrams constructed in this work. In figure~\ref{fig:fig4}, the diagram $(\mu, \mu')$ is presented. As one can see the general form and coordinates of characteristic points [on the left-hand side (hard-core bosons) and first part on the right-hand side (the case of three-state model)] are close to each other. 
In figure~\ref{fig:fig5}, the diagrams $(\mu, |t_0|)$ are presented. Here again we can see the similarity between the figure on the left-hand side and the related part (in the vicinity of the value $\mu=\mu'$) on the right-hand side diagram for these two cases. 

\section{Conclusions}

In the cases where transition to the SF phase is accompanied by the change of the mean number of fermions, the PT order changes from the 2nd to the 1st one; this result corresponds to the literature data (DMRG calculations for harmonic trap, \cite{snoek:155301}).

The shape of the MI phase regions (so-called lobes) in the $(\mu,t_0)$ diagrams and the localization of the 1st order PT segments upon them depend on the $\mu'$ level position.

The appearance of fermions  worsens the conditions for the BE-condensate existence on the background of unoccupied fermion states.
At the same time, the existence of fermions stimulates the appearance of the BE-condensate of the other type \cite{stasyuk:96}. For example, in \cite{gunter:180402} and \cite{ospelkaus:120402}, the mixture of fermionic $^{40}\text{K}$ and bosonic $^{87}\text{Rb}$ atoms in a three-dimensional optical lattice was studied and it was observed that an increasing admixture of the fermionic species diminishes the phase coherence of the bosonic atoms. In \cite{best:030408}, for the same set of fermionic and bosonic atoms ($^{40}\text{K}$ and $^{87}\text{Rb}$), there was noticed a shift of the line of MI-SF phase transition depending on the sign and absolute value of boson-fermion interaction (which can be tuned in a wide range  using Feshbach resonance). 

 Consideration in the hard-core boson limit allows one to describe SF-MI transitions in the regions of the MI lobes contact points. The $n^\text{b} \leqslant 2$ restriction enables one to analyze the more general case of separated lobes. That is enough to understand the structure of $(\mu, |t_0|)$ diagrams at arbitrary $n_\text{B}$ values with an account of the possibility of the 1st order PT.



%
%

\ukrainianpart

\title{Модель Бозе-Фермі-Хаббарда в обмеженому гільбертовому просторі}
\author{В.О. Краснов, \framebox{І.В. Стасюк}}
\address{
	Iнститут фiзики конденсованих систем НАН України, вул. Свєнцiцького, 1, 79011 Львiв, Україна
}
%
%
%

\makeukrtitle

\begin{abstract}
\tolerance=3000%
Показано, що в моделі Бозе-Фермі-Хаббарда, що використовується для опису надхолодних атомарних бозе-фермі сумішей в оптичних гратках, обмеження  $n_\text{B} \leqslant 2$ дозволяє проаналізувати більш загальні випадки розділених куполоподібних областей існування фази моттівського діелектрика в порівнянні з випадком $n_\text{B} \leqslant 1$ (наближення жорстких бозонів).  Також показано, що обмеження за числом бозонів не більшим ніж два на один вузол, є достатньо  для того щоб зрозуміти структуру $(\mu, |t_0|)$ діаграм при довільному значенні $n^\text{b}$ з врахуванням можливості існування фазових пепреходів 1-го роду.
 
\keywords модель Бозе-Фермі-Хаббарда, бозе-конденсат, оптичні гратки, фазові переходи

\end{abstract}

\end{document}